\documentclass[11pt]{article}

\usepackage[margin=1in]{geometry}
\usepackage[authoryear]{natbib}
\usepackage{xcolor}
\usepackage[utf8]{inputenc}
\usepackage[T1]{fontenc}    
\usepackage{hyperref}       
\usepackage{url}            
\usepackage{booktabs}       
\usepackage{amsfonts}      
\usepackage{nicefrac}       
\usepackage{microtype}     
\usepackage{enumitem}
\usepackage{amsmath}
\usepackage{amsthm}
\usepackage{bbm}
\usepackage{graphicx}
\usepackage{float}
\usepackage{subcaption}
\usepackage{datetime2}
\usepackage{setspace}
\usepackage{rotating}
\usepackage{comment}
\usepackage{listings}
\usepackage{multirow}
\usepackage{tcolorbox}
\usepackage{tikz}
\usepackage{comment}
\usetikzlibrary{arrows.meta, decorations.pathreplacing, calc}

\hypersetup{
    bookmarks=false,                 	% show bookmarks bar?
    unicode=false,                  	% non-Latin characters in bookmarks
    pdftoolbar=true,                	% show toolbar?
    pdfmenubar=true,                	% show menu?
    pdffitwindow=true,              	% window fit to page when opened
    pdfstartview={FitH},            	% fits the width of the page to the window
    pdftitle={},                    		% title
    pdfauthor={Ryan Kessler},	% author
    pdfnewwindow=true,              % links in new window
    colorlinks=true,                	% false: boxed links; true: colored links
    linkcolor=blue,                 	% color of internal links
    citecolor=blue,                 	% color of links to bibliography
    filecolor=magenta,              	% color of file links
    urlcolor=blue,                  	% color of external links
    breaklinks=true
}

\DeclareMathOperator*{\argmax}{arg\,max}

\title{Robust Estimation of Prior Parameters in Normal-Normal Empirical Bayes}

\author{
  Manu Burson\\
  \texttt{msburson@amazon.com}
  \and
  Ryan Kessler\\
  \texttt{rykessle@amazon.com}
}
\date{\today}

\begin{document}

\maketitle

\begin{abstract}
Many large-scale A/B testing systems leverage normal-normal empirical Bayes models to improve estimates of treatment impacts and launch decisions. We show that in these models maximum likelihood estimates of the variance of true impacts can be sensitive to a small number of extreme observations. We propose a bounded influence maximum likelihood estimator (BIMLE) that downweights such observations in proportion to their inconsistency with the model, reducing sensitivity while retaining the full sample in estimation. Absent extreme observations, BIMLE coincides with the standard MLE. In an application to A/B tests run at Amazon, BIMLE reduces instability in the estimated prior variance by over 50 percent while downweighting less than 0.1 percent of observations. 
\end{abstract}

\section{Introduction}

Many large-scale A/B testing systems leverage normal-normal empirical Bayes to improve estimates of treatment impacts and launch decisions \citep{eppo_2025, statsig_2025, sudijono_etal_2024, mudd_etal_2026}. Statistical power in A/B tests is often limited, increasing the risk that launch decisions are driven by noise rather than true impact \citep{kohavi_etal_2024, gelman_carlin_2014}. By pooling information across tests, empirical Bayes methods can reduce estimation error and improve decision making \citep{efron_morris_1975}. The normal-normal model is particularly attractive because it combines these statistical benefits with simple, closed-form shrinkage rules that are easy to implement at scale in production systems. 

This requires estimating the mean and variance of the distribution of true impacts. Because true impacts are unobserved, these parameters must be inferred from noisy estimates of impacts. In practice, this is often done by pooling information across large samples of tests and estimating the parameters via maximum likelihood \citep{azevedo_etal_2019}. The resulting estimates inform shrinkage rules that are applied across the A/B testing system, often influencing the estimated impacts and launch decisions of thousands of tests. 

In this paper, we show that estimates of these parameters can be sensitive to the results of just a few tests, making it difficult to leverage them in production systems at scale. This sensitivity arises because ``extreme'' test results can exert effectively unbounded influence on the estimated prior. A natural response is to identify and exclude such tests from the estimation sample. However, ``extreme'' test results can reflect either genuinely large impacts or data quality issues, and distinguishing between these possibilities is often difficult in large-scale A/B testing systems. The choice of which tests to exclude is therefore necessarily ad hoc, and can itself shift the resulting estimates substantially. Even when problematic tests can be identified ex post, new failure modes continually emerge, so maintaining perfect screening is rarely feasible in practice. 

We introduce a bounded influence maximum likelihood estimator (BIMLE) of the parameters to reduce this sensitivity. Adapting ideas from the robust statistics literature \citep{huber_1964, hampel_1974, hampel_etal_1986}, BIMLE retains all observations in estimation but downweights tests in proportion to how far their results are from the model's predictions, thereby limiting the influence that any individual test can exert on the estimated prior. The estimator is governed by a tuning constant $b$, with lower values providing greater protection against ``extreme'' test results. As $b \to \infty$ all weights converge to one and BIMLE reduces to the standard maximum likelihood estimator (MLE). BIMLE therefore nests the MLE as a special case, while providing a practical mechanism for controlling sensitivity to the results of individual tests.  

In simulations and an application to A/B tests run at Amazon, we find that BIMLE can improve the stability of prior estimates while downweighting only a small fraction of observations. In simulations, we show that replacing just 5 of 1K test results with extreme values can inflate the MLE of prior variance by more than tenfold, while BIMLE increases only modestly. Alternative estimators --- such as maximum a posteriori (MAP) estimation with correctly centered, informative hyperpriors --- provide little additional protection. In an application to thousands of A/B tests run at Amazon, MLE estimates of the prior standard deviation swing by up to an order of magnitude across adjacent half-year periods, while BIMLE is stable and agrees with the MLE in periods without ``extreme'' test results. These gains come at minimal cost, with BIMLE downweighting fewer than 0.1 percent of observations under our recommended tuning of $b = 15$. 

Our main contribution is practical, not theoretical. We highlight a failure mode in the estimation of the normal-normal empirical Bayes model in large-scale A/B testing systems. We then show how ideas from the robust statistics literature can be adapted to address the problem, while retaining the simplicity and scalability of the model. We view this work as part of a larger effort to enable Bayesian methods at scale responsibly. Prior work has focused on making estimates of the posterior mean robust to misspecification in the prior \citep{kessler_2025}. In this paper, we focus on the complementary problem of how to make estimates of the prior robust to ``extreme'' test results. Together, these approaches may help large-scale A/B testing systems realize the benefits of Bayesian methods while accounting for imperfect data and misspecification that inevitably arise in practice. 

\section{Model and Estimation}

\subsection{Setup}
\label{sec:setup}

We consider a firm that seeks to construct an estimator of the true impact $\Delta_i$ of A/B test $i$ that performs well on average across $i \in \{1,\ldots,N\}$ independent tests. For each test, the firm observes a realization of an unbiased estimator $\hat{\Delta}_i$, with known sampling variance $\tau_i^2$. By virtue of the central limit theorem, we assume: 

\begin{equation}
    \hat\Delta_i | \Delta_i, \tau_i^2 \sim N(\Delta_i, \tau_i^2) \label{eq:sampling}
\end{equation}

Because the estimator $\hat{\Delta}_i$ is noisy --- with true impacts $\Delta_i$ small in magnitude relative to $\tau_i^2$ --- the firm combines information across tests to improve estimation accuracy. As a working model, the firm assumes that the true impacts are normally distributed across tests, with unknown mean $\mu$ and variance $\sigma^2$:

\begin{equation}
    \Delta_i | \mu, \sigma^2 \sim N(\mu, \sigma^2) \label{eq:prior}
\end{equation}

The firm then considers the following plug-in estimator of $\Delta_i$:

\begin{equation}
    m\left(\hat{\mu}, \hat{\sigma}^2\right) = \left(\frac{\hat{\sigma}^2}{\hat{\sigma}^2 + \tau_i^2}\right) \hat{\Delta}_i + \left(\frac{\tau_i^2}{\hat{\sigma}^2 + \tau_i^2}\right) \hat{\mu} \label{eq:posterior_mean}
\end{equation}

where $(\hat{\mu}, \hat{\sigma}^2)$ are estimates of $(\mu, \sigma^2)$ obtained by pooling data across tests.

This estimator $m\left(\hat{\mu}, \hat{\sigma}^2\right)$ minimizes mean-squared error (MSE) across tests under two distinct assumptions on the prior. First, if the prior is normal, as in equation (\ref{eq:prior}), the estimator is the posterior mean of $\Delta_i$ given $(\hat{\Delta}_i, \tau_i^2, \mu, \sigma^2)$ and therefore minimizes MSE \citep{casella_berger_2002}. Second, even if the prior is not normal, the estimator still performs well whenever $E(\Delta_i) = \mu$ and $\text{Var}(\Delta_i) = \sigma^2$. Among estimators linear in $\hat\Delta_i$, it minimizes MSE for any such prior \citep{berger_1985}. Allowing for estimators that are nonlinear in $\hat\Delta_i$, it minimizes the worst-case MSE across all such priors \citep{berger_1985}. The normality assumption in equation (\ref{eq:prior}) is therefore a convenience for motivating $m(\hat\mu, \hat\sigma^2)$, not a necessity for it to perform well.

\subsection{Maximum Likelihood Estimation}
\label{sec:mle}

To operationalize the estimator $m\left(\hat{\mu}, \hat{\sigma}^2\right)$, the firm estimates the parameters $(\mu, \sigma^2)$ via maximum likelihood. Under equations (\ref{eq:sampling}) and (\ref{eq:prior}), the marginal distribution of $\hat{\Delta}_i$ (unconditional on $\Delta_i$) is normal, with mean $\mu$ and variance $\sigma^2 + \tau_i^2$: 

\begin{equation}
    \hat{\Delta}_i | \mu, \sigma^2 \sim N\left(\mu, \sigma^2 + \tau_i^2\right)
\end{equation}

The log likelihood of observing $\{\hat{\Delta}_i, \tau_i^2\}_{i=1}^N$ given $(\mu, \sigma^2)$ is therefore: 

\begin{equation}
    \mathcal{L}(\mu, \sigma^2) = \sum_{i=1}^N \mathcal{L}_i(\mu, \sigma^2) = -\frac{1}{2}\sum_{i=1}^N \left[\log(\sigma^2 + \tau_i^2) + \frac{(\hat\Delta_i - \mu)^2}{\sigma^2 + \tau_i^2}\right] \label{eq:loglik}
\end{equation}

This motivates the maximum likelihood estimator (MLE):

\begin{equation}
    \left(\hat{\mu}_{M}, \hat{\sigma}_{M}^2\right) = \argmax_{\mu, \sigma^2 \geq 0} \mathcal{L}(\mu, \sigma^2) \label{eq:mle}
\end{equation}

Under standard regularity conditions, the MLE is consistent and asymptotically normal. Consistency holds even if the true impacts are not normally distributed, as assumed in equation (\ref{eq:prior}) \citep{gourieroux_etal_1984}.

\section{The Problem: Influence of Extreme Observations}

The MLEs $(\hat{\mu}_M, \hat{\sigma}_M^2)$ can be driven by a small number of ``extreme'' observations, leading to estimates that are sensitive to the particular sample of tests. Because the MLEs are used in estimating the impact of all tests, this means a small number of tests can distort launch decisions across the firm.

The first-order conditions underlying equation (\ref{eq:mle}) show why. The MLE requires that the scores (partial derivatives of $\mathcal{L}$) equal zero at the estimated parameter values: 

\vspace{-1pc}
\begin{align}
    \frac{\partial \mathcal{L}}{\partial \mu}\bigg|_{(\hat{\mu}_M, \hat{\sigma}_M^2)} &= \sum_{i=1}^N s_i(\mu)\bigg|_{(\hat{\mu}_M, \hat{\sigma}_M^2)} = \sum_{i=1}^N \frac{z_i}{\sqrt{\hat{\sigma}_M^2 + \tau_i^2}} = 0 \label{eq:foc_mu} \\
    \frac{\partial \mathcal{L}}{\partial \sigma^2}\bigg|_{(\hat{\mu}_M, \hat{\sigma}_M^2)} &= \sum_{i=1}^N s_i(\sigma^2)\bigg|_{(\hat{\mu}_M, \hat{\sigma}_M^2)} = \sum_{i = 1}^N \frac{z_i^2 - 1}{2(\hat{\sigma}_M^2 + \tau_i^2)} = 0 \label{eq:foc_sigma}
\end{align}

where $z_i = (\hat\Delta_i - \hat{\mu}_M) / \sqrt{\hat{\sigma}_M^2 + \tau_i^2}$. When a test is added to the sample, the MLE must adjust $(\hat{\mu}_M,\hat{\sigma}_M^2)$ until both scores return to zero. The amount the test influences the estimates therefore depends on how much it contributes to each score \citep{hampel_1974}. The scores in equations (\ref{eq:foc_mu}) and (\ref{eq:foc_sigma}) exhibit an important asymmetry. The score for $\mu$ is linear in $z_i$, while the score for $\sigma^2$ is quadratic in $z_i$. As a result, tests with large $|z_i|$ contribute disproportionately to the estimation of $\sigma^2$, allowing a small number of tests to drive the estimate $\hat{\sigma}_M^2$. 

This sensitivity is particularly problematic when the firm is testing at scale. Extreme test results can reflect genuinely large effects or data quality issues (such as randomization failure or metric misconfiguration) and the firm often lacks the resources and domain context to distinguish between the two. Robust estimation therefore requires limiting the influence any single test has on the scores in equations (\ref{eq:foc_mu}) and (\ref{eq:foc_sigma}).

\section{The Solution: Bounded Influence Maximum Likelihood Estimation}

We introduce a bounded influence maximum likelihood estimator (BIMLE) that uses the full sample of tests while limiting the influence of any given test \citep{huber_1964, hampel_1974, field_smith_1994}.

BIMLE bounds each test's influence via weighted maximum likelihood estimation, with weights reflecting the plausibility of each test's results under the model. Under equations (\ref{eq:sampling}) and (\ref{eq:prior}), $z_i \sim N(0,1)$, so tests with large $|z_i|$ cast doubt on the model. From equation (\ref{eq:foc_sigma}), these same tests often dominate the score for $\sigma^2$ and thereby shape the estimate $\hat{\sigma}_M^2$. We bound each test's score contribution by specifying weights:

\begin{equation}
w_i(\mu, \sigma^2; b) = \min\left(1, \frac{b^2}{z_i^2}\right),
\end{equation}

where $b \geq 0$ is a firm choice parameter. Tests with $|z_i| \leq b$ are more consistent with the model and receive a weight of 1. Tests with $|z_i| > b$ are less consistent with the model and are therefore downweighted, with the weight decreasing in $z_i^2$ so that the weighted contribution $w_i \cdot (z_i^2 - 1)$ to the score for $\sigma^2$ is bounded above by $b^2$.

The weights are based on $z_i$ rather than the test statistic $t_i = \hat\Delta_i / \tau_i$ because whether a test result is ``extreme'' depends on $\sigma^2$, which is itself one of the parameters being estimated. Consider two metrics for a test with $\tau_i^2 = 1$. Metric A has large prior variance ($\sigma^2 = 4$), so $\text{Var}(t_i) = 5$ and $|t_i| = 5$ is well within normal variation. Metric B has small prior variance ($\sigma^2 = 0.1$), so $\text{Var}(t_i) = 1.1$ and the same $|t_i| = 5$ is highly unusual. A fixed $t$-threshold is therefore a blunt instrument. When it is set at 5, legitimate results for Metric A are incorrectly excluded. When it is set at 10, problematic results for Metric B are incorrectly included. The right threshold depends on $\sigma^2$, which is unknown. Weighting by $z_i$, which standardizes by $\sqrt{\sigma^2 + \tau_i^2}$ rather than $\tau_i$ alone, addresses this problem.

Given $b \geq 0$, the BIMLE $(\hat{\mu}_B, \hat{\sigma}_B^2)$ is defined as the fixed point of the weighted score equations:

\vspace{-1pc}
\begin{align}
    \sum_{i=1}^N w_i(\mu, \sigma^2; b) \, s_i(\mu) \bigg|_{(\hat{\mu}_B, \hat{\sigma}_B^2)} &= 0 \label{eq:bimle_mu} \\
    \sum_{i=1}^N w_i(\mu, \sigma^2; b) \, s_i(\sigma^2) \bigg|_{(\hat{\mu}_B, \hat{\sigma}_B^2)} &= 0 \label{eq:bimle_sigma}
\end{align}

Because the weights depend on $(\mu, \sigma^2)$ through $z_i$, the estimator is defined implicitly. We solve it via iteratively reweighted maximum likelihood:

\begin{enumerate}
\item Initialize: estimate $\left(\hat\mu_B^{(0)}, \hat\sigma_B^{2(0)}\right)$ via MLE on the full sample.

\item Compute residuals: $z_i^{(t)} = \left(\hat\Delta_i - \hat\mu_B^{(t)}\right) / \sqrt{\hat\sigma_B^{2(t)} + \tau_i^2}$.

\item Update weights: $w_i^{(t)} = \min\left(1,\; b^2 / z_i^{2(t)}\right)$.

\item Re-estimate: holding the weights fixed at $w_i^{(t)}$, maximize $\sum_{i=1}^N w_i^{(t)} \, \mathcal{L}_i\left(\mu, \sigma^2\right)$ to obtain $\left(\hat\mu_B^{(t+1)}, \hat\sigma_B^{2(t+1)}\right)$.

\item Repeat steps 2--4 until convergence.
\end{enumerate}

In practice, convergence typically occurs in fewer than 10 iterations.

The estimator has four attractive properties. First, no single test's contribution to the score equations (\ref{eq:bimle_mu})--(\ref{eq:bimle_sigma}) exceeds $b^2$, directly addressing the unbounded influence of the MLE. Second, with a small bias correction to account for the systematic downweighting of tail observations \citep{field_smith_1994}, the estimator is consistent under the model.\footnote{The bias correction is of order $\text{Pr}(|z_i| > b)$. At the recommended $b = 15$, $\text{Pr}(|z_i| > 15) < 10^{-10}$ and we omit it in practice.} Third, it lets the firm trade responsiveness (high $b$) against robustness (low $b$) through the single constant $b$. As $b \to \infty$ all weights converge to one and BIMLE reduces to the standard MLE. Fourth, it complements the robust Bayes estimator of \citet{kessler_2025}, who uses the same $z_i$ to determine the weight on the prior when estimating $\Delta_i$. Together, the two bound influence at both stages of normal-normal empirical Bayes: BIMLE bounds the influence of an anomalous test on the estimated prior, and \citet{kessler_2025} bounds its influence on the estimate of its own effect $\Delta_i$ given the prior.

The primary limitation of the estimator is that it requires the firm to choose $b$. The choice reflects a tradeoff between robustness and responsiveness: smaller values provide greater protection against data quality issues but may downweight legitimately large impacts, while larger values retain sensitivity to large impacts but provide less protection. We recommend a default of $b = 15$. Under the model, observations with $|z_i| > 15$ are unlikely --- with $\Pr(|z_i| > 15) < 10^{-10}$ --- so BIMLE intervenes only on tests clearly inconsistent with the model. In our empirical application, this downweights fewer than 0.1 percent of observations while substantially improving stability. Moreover, results are not sensitive to this choice. Appendix \ref{sec:appendix_supplemental} shows that varying $b$ across $\{10, 15, 20\}$ shifts the magnitude of $\hat\sigma$ modestly while the coefficient of variation remains low for all choices.

A second limitation is that calibrating $b$ leans on the normality assumption in equation (\ref{eq:prior}). The MLE $(\hat\mu_M, \hat\sigma_M^2)$ remains consistent and $m(\hat\mu, \hat\sigma^2)$ continues to perform well even when the true impacts are not normally distributed, so long as their mean and variance are $\mu$ and $\sigma^2$ (sections \ref{sec:mle} and \ref{sec:setup}). But calibrating $b$ against a tail probability requires more than knowing these two moments. It requires the full distribution of $z_i$ under the model. If true impacts are not normally distributed, this calibration will no longer hold exactly. 

\section{Shortcomings of Alternative Approaches}

An alternative approach is to estimate $(\mu, \sigma^2)$ via their posterior mode (MAP), with hyperpriors on $(\mu, \sigma^2)$. By encoding beliefs about the plausible range of $(\mu, \sigma^2)$, hyperpriors can in principle prevent the estimates from being driven by extreme observations. Assuming independent hyperprior densities $\pi(\mu)$ and $\pi(\sigma^2)$, Bayes' rule gives a posterior $p(\mu, \sigma^2 | \text{data}) \propto L(\text{data} | \mu, \sigma^2) \cdot \pi(\mu) \cdot \pi(\sigma^2)$. The MAP estimator maximizes the log posterior, or equivalently, the penalized log likelihood:

\begin{equation}
    \left(\hat{\mu}_{P}, \hat{\sigma}_{P}^2\right) = \argmax_{\mu, \sigma^2 \geq 0} \left[\mathcal{L}(\mu, \sigma^2) + \log \pi(\mu) + \log \pi(\sigma^2)\right] \label{eq:map}
\end{equation}

The MAP estimator generally does not solve the sensitivity problem. The score for $\sigma^2$ is:

\begin{equation}
    \underbrace{\sum_{i=1}^N \frac{z_i^2 - 1}{2(\sigma^2 + \tau_i^2)}}_{\text{MLE score (unbounded)}} + \underbrace{\frac{\partial \log \pi(\sigma^2)}{\partial \sigma^2}}_{\text{hyperprior (bounded)}} = 0 \label{eq:map_foc}
\end{equation}

The hyperprior contributes the term $\partial \log \pi(\sigma^2) / \partial \sigma^2$, which depends only on $\sigma^2$ and is bounded regardless of the data, whereas each observation's contribution to the first sum grows with $z_i^2$ and is unbounded. For any hyperprior, sufficiently ``extreme'' observations therefore dominate the score and pull $\hat{\sigma}_P$ upward. In other words, hyperpriors shift the score but they do not bound it, leaving the fundamental problem identified in section 3 intact. 

To help see this, consider the log-normal hyperprior $\log(\sigma^2) \sim N(\eta, \zeta^2)$ for which:

\begin{equation}
   \lim_{\sigma^2 \rightarrow \infty} \left( \frac{\partial \log \pi(\sigma^2)}{\partial \sigma^2}\right)
    = \lim_{\sigma^2 \rightarrow \infty} \left( -\frac{1}{\sigma^2}\left(1 + \frac{\log \sigma^2 - \eta}{\zeta^2}\right)\right)  = 0
\end{equation}

The hyperprior exerts downward pressure on the estimate of $\sigma^2$ when $\log \sigma^2 > \eta - \zeta^2$ and therefore $\partial \log \pi(\sigma^2) / \partial \sigma^2 < 0$. But its strength vanishes as $\sigma^2$ grows. The hyperprior therefore loses its regularizing power precisely when extreme observations have already pushed the estimate into the region where regularization is needed.

Heavy-tailed priors do not solve the sensitivity problem either. A small number of observations can dominate estimation of the mean and variance of true impacts regardless of the assumed prior family. Heavy-tailed priors also introduce additional parameters to be estimated (e.g., degrees of freedom) and sacrifice the closed-form shrinkage rule of equation (\ref{eq:posterior_mean}), making them more difficult to deploy at scale.\footnote{\citet{mudd_etal_2026} show that a normal-inverse-gamma mixture can preserve conjugacy while allowing for heavier-than-normal tails, avoiding this tradeoff. But this approach still requires estimating the parameters of a normal marginal likelihood, which remains susceptible to the unbounded influence of extreme observations we highlight.} BIMLE instead keeps the normal-normal model intact and addresses the source of sensitivity directly.

\section{Illustrative Simulations}
\label{sec:simulation}

We compare the MLE, BIMLE, and MAP estimator via a simple contamination experiment. We draw $N = 1{,}000$ tests from the model with $\mu = 0$, $\sigma^2 = 1$, and $\tau_i^2 = 1$ for all $i$, then replace $k \in \{0, 1, 2, 3, 4, 5\}$ observations with $\hat{\Delta}_i = 50$. For each sample, we compute the MLE, BIMLE with $b = 10$, and the MAP estimator from equation (\ref{eq:map}) with $\pi(\mu) = N(0, 1)$ and $\pi(\sigma^2)$ implied by $\log(\sigma^2) \sim N(0, 0.25^2)$.\footnote{We use $b = 10$ here only to illustrate the mechanism for BIMLE in a small sample; we use the recommended $b = 15$ elsewhere.}  Both hyperpriors are informative and correctly centered at the true parameter values (95\% of the mass on $\mu \in [-2, 2]$ and $\sigma^2 \in [0.6, 1.6]$). We repeat this exercise 100 times and report averages across replications. 

\begin{figure}[ht]
    \centering
    \caption{Simulation results: Sensitivity of estimators to ``extreme'' observations}
    \label{fig:influence_comparison}
    \includegraphics[width=\textwidth]{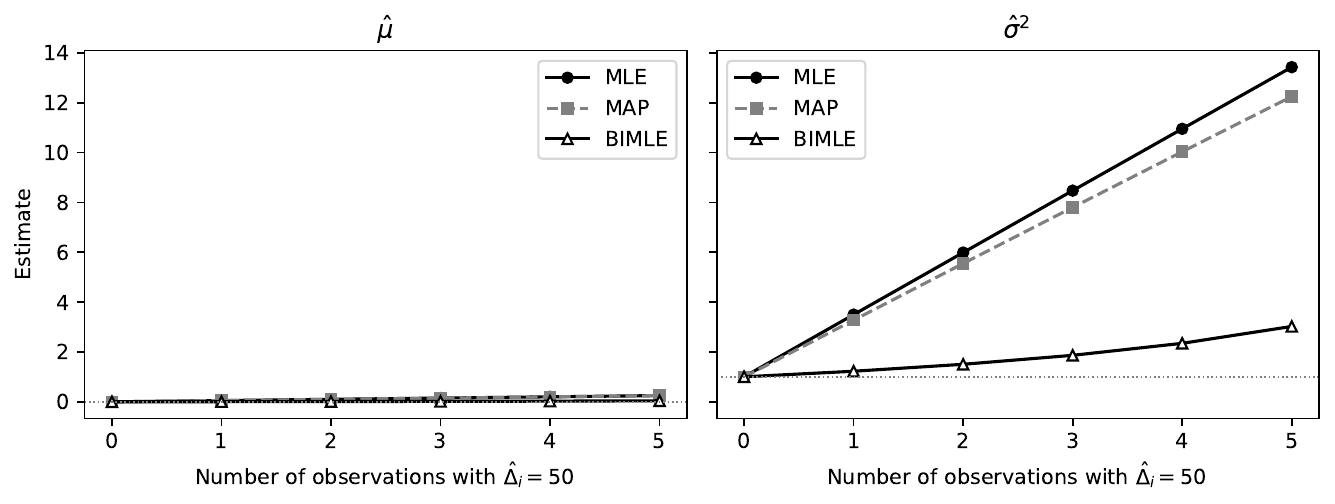} 
\end{figure}

Figure \ref{fig:influence_comparison} reports the results. When there are no extreme observations ($k = 0$), all three estimators recover $\mu = 0$ and $\sigma^2 = 1$, confirming that BIMLE does not distort estimation under the model. As the number of extreme observations increases, the MLE becomes highly sensitive: replacing one or two observations (0.1–0.2 percent of the sample) substantially increases the estimated prior variance, with $\hat{\sigma}_M^2$ rising more than tenfold as $k$ increases from 0 to 5. The MAP estimator offers limited protection because the information from extreme observations quickly overwhelms the hyperprior regularization. In contrast, BIMLE remains stable. The estimated prior variance increases from about 1.0 to 3.0 as $k$ rises from 0 to 5, indicating that BIMLE remains responsive to potentially informative observations while preventing a few tests from determining the estimate. The left panel shows that all three estimators are much less sensitive when estimating $\mu$, consistent with the score for $\mu$ being linear rather than quadratic in $z_i$.

\section{Empirical Application}
\label{sec:empirical}

We apply the MLE, MAP, and BIMLE to results from A/B tests run at Amazon, focusing on four illustrative metrics. For each test, we restrict attention to results as of the most recent start date and longest duration, and we exclude tests with known allocation issues. We estimate $(\mu, \sigma^2)$ separately for each metric in each half-year period, yielding six independent estimates per metric.

\begin{table}[ht]
\centering
\caption{Estimates of $\hat\sigma \times 100$ by half-year period}
\label{tab:empirical_stability}
\begin{subtable}[t]{0.48\textwidth}
\centering
\caption{Metric A}
{\scriptsize
\renewcommand{\arraystretch}{0.85}
\begin{tabular}{lrrr}
\toprule
 & \multicolumn{1}{c}{MLE} & \multicolumn{1}{c}{MAP} & \multicolumn{1}{c}{BIMLE} \\
\midrule
2023-H1 & \shortstack{0.016\\{}[0.008, 0.021]} & \shortstack{0.015\\{}[0.007, 0.021]} & \shortstack{0.016\\{}[0.008, 0.021]} \\
2023-H2 & \shortstack{0.019\\{}[0.014, 0.024]} & \shortstack{0.019\\{}[0.013, 0.024]} & \shortstack{0.019\\{}[0.014, 0.024]} \\
2024-H1 & \shortstack{0.021\\{}[0.012, 0.031]} & \shortstack{0.021\\{}[0.011, 0.030]} & \shortstack{0.021\\{}[0.011, 0.030]} \\
2024-H2 & \shortstack{0.071\\{}[0.028, 0.641]} & \shortstack{0.070\\{}[0.028, 0.621]} & \shortstack{0.026\\{}[0.019, 0.034]} \\
2025-H1 & \shortstack{0.021\\{}[0.013, 0.031]} & \shortstack{0.021\\{}[0.012, 0.031]} & \shortstack{0.020\\{}[0.012, 0.029]} \\
2025-H2 & \shortstack{0.900\\{}[0.389, 1.201]} & \shortstack{0.900\\{}[0.389, 1.201]} & \shortstack{0.023\\{}[0.016, 0.030]} \\
\midrule
CV & 1.86 & 1.86 & 0.15 \\
DW & -- & -- & 0.02\% \\
$w | $ DW & -- & -- & 0.50 \\
\bottomrule
\end{tabular}
}
\end{subtable}
\hfill
\begin{subtable}[t]{0.48\textwidth}
\centering
\caption{Metric B}
{\scriptsize
\renewcommand{\arraystretch}{0.85}
\begin{tabular}{lrrr}
\toprule
 & \multicolumn{1}{c}{MLE} & \multicolumn{1}{c}{MAP} & \multicolumn{1}{c}{BIMLE} \\
\midrule
2023-H1 & \shortstack{0.031\\{}[0.023, 0.036]} & \shortstack{0.030\\{}[0.022, 0.036]} & \shortstack{0.031\\{}[0.023, 0.036]} \\
2023-H2 & \shortstack{0.022\\{}[0.013, 0.028]} & \shortstack{0.021\\{}[0.011, 0.028]} & \shortstack{0.022\\{}[0.013, 0.028]} \\
2024-H1 & \shortstack{0.032\\{}[0.017, 0.045]} & \shortstack{0.031\\{}[0.016, 0.045]} & \shortstack{0.031\\{}[0.017, 0.043]} \\
2024-H2 & \shortstack{0.029\\{}[0.020, 0.037]} & \shortstack{0.029\\{}[0.019, 0.037]} & \shortstack{0.029\\{}[0.019, 0.037]} \\
2025-H1 & \shortstack{0.021\\{}[0.011, 0.029]} & \shortstack{0.020\\{}[0.009, 0.028]} & \shortstack{0.021\\{}[0.011, 0.029]} \\
2025-H2 & \shortstack{0.849\\{}[0.035, 1.290]} & \shortstack{0.849\\{}[0.035, 1.290]} & \shortstack{0.033\\{}[0.023, 0.047]} \\
\midrule
CV & 1.87 & 1.88 & 0.17 \\
DW & -- & -- & 0.01\% \\
$w | $ DW & -- & -- & 0.41 \\
\bottomrule
\end{tabular}
}
\end{subtable}

\begin{subtable}[t]{0.48\textwidth}
\centering
\caption{Metric C}
{\scriptsize
\renewcommand{\arraystretch}{0.85}
\begin{tabular}{lrrr}
\toprule
 & \multicolumn{1}{c}{MLE} & \multicolumn{1}{c}{MAP} & \multicolumn{1}{c}{BIMLE} \\
\midrule
2023-H1 & \shortstack{0.292\\{}[0.221, 0.366]} & \shortstack{0.292\\{}[0.221, 0.366]} & \shortstack{0.209\\{}[0.187, 0.230]} \\
2023-H2 & \shortstack{0.660\\{}[0.254, 0.998]} & \shortstack{0.660\\{}[0.254, 0.998]} & \shortstack{0.232\\{}[0.208, 0.259]} \\
2024-H1 & \shortstack{1.268\\{}[0.600, 1.789]} & \shortstack{1.268\\{}[0.600, 1.789]} & \shortstack{0.377\\{}[0.305, 0.440]} \\
2024-H2 & \shortstack{0.926\\{}[0.684, 1.155]} & \shortstack{0.926\\{}[0.684, 1.155]} & \shortstack{0.305\\{}[0.272, 0.336]} \\
2025-H1 & \shortstack{0.361\\{}[0.242, 0.503]} & \shortstack{0.361\\{}[0.242, 0.502]} & \shortstack{0.209\\{}[0.182, 0.237]} \\
2025-H2 & \shortstack{1.423\\{}[0.973, 1.730]} & \shortstack{1.423\\{}[0.972, 1.730]} & \shortstack{0.238\\{}[0.205, 0.276]} \\
\midrule
CV & 0.52 & 0.52 & 0.23 \\
DW & -- & -- & 0.07\% \\
$w | $ DW & -- & -- & 0.40 \\
\bottomrule
\end{tabular}
}
\end{subtable}
\hfill
\begin{subtable}[t]{0.48\textwidth}
\centering
\caption{Metric D}
{\scriptsize
\renewcommand{\arraystretch}{0.85}
\begin{tabular}{lrrr}
\toprule
 & \multicolumn{1}{c}{MLE} & \multicolumn{1}{c}{MAP} & \multicolumn{1}{c}{BIMLE} \\
\midrule
2023-H1 & \shortstack{0.113\\{}[0.064, 0.300]} & \shortstack{0.113\\{}[0.064, 0.300]} & \shortstack{0.071\\{}[0.058, 0.081]} \\
2023-H2 & \shortstack{0.318\\{}[0.075, 0.583]} & \shortstack{0.317\\{}[0.075, 0.583]} & \shortstack{0.078\\{}[0.066, 0.089]} \\
2024-H1 & \shortstack{0.127\\{}[0.085, 0.194]} & \shortstack{0.127\\{}[0.085, 0.194]} & \shortstack{0.090\\{}[0.073, 0.107]} \\
2024-H2 & \shortstack{0.550\\{}[0.124, 0.914]} & \shortstack{0.550\\{}[0.124, 0.914]} & \shortstack{0.102\\{}[0.081, 0.121]} \\
2025-H1 & \shortstack{0.068\\{}[0.054, 0.080]} & \shortstack{0.068\\{}[0.054, 0.080]} & \shortstack{0.066\\{}[0.053, 0.078]} \\
2025-H2 & \shortstack{0.992\\{}[0.302, 1.395]} & \shortstack{0.992\\{}[0.301, 1.395]} & \shortstack{0.073\\{}[0.062, 0.085]} \\
\midrule
CV & 0.90 & 0.90 & 0.15 \\
DW & -- & -- & 0.04\% \\
$w | $ DW & -- & -- & 0.37 \\
\bottomrule
\end{tabular}
}
\end{subtable}
\vspace{0.1pc}
{\scriptsize
\begin{flushleft}
Notes: BIMLE uses $b = 15$. MAP uses $\mu \sim N(0, 10^{-6})$ and $\log(\sigma^2) \sim N(-16, 2^2)$. Brackets show 95\% confidence intervals from a nonparametric bootstrap (200 resamples of the tests within each half-year window). CV is the coefficient of variation across half-year periods. DW is the average percent of observations receiving weight less than 1. $w | $ DW is the average weight among downweighted observations.
\end{flushleft}}
\end{table}

Table \ref{tab:empirical_stability} reports $\hat\sigma \times 100$ for four metrics, along with the coefficient of variation (CV = standard deviation / mean) of estimates across periods. The MLE is unstable. For Metric A, the estimated prior standard deviation ranges from 0.016 to 0.900 across periods (CV = 1.86). For Metric C, it ranges from 0.292 to 1.423 (CV = 0.52). The MAP estimator --- computed with informative hyperpriors $\mu \sim N(0, 10^{-6})$ (centered at zero with 95\% of the mass on $|\mu| \leq 0.2\%$) and $\log(\sigma^2) \sim N(-16, 2^2)$ (centered at $\sigma = 0.03\%$ with 95\% of the mass between 0.005\% and 0.24\%) --- provides no meaningful regularization. For every metric, the MAP estimates are nearly identical to the MLE, confirming that the likelihood from extreme observations overwhelms even a correctly centered hyperprior.

BIMLE with $b = 15$ produces stable estimates across all metrics and periods. For Metric A, the estimated prior standard deviation ranges from 0.016 to 0.026 (CV = 0.15). For Metric C, it ranges from 0.209 to 0.377 (CV = 0.23). This stability is achieved by downweighting fewer than 0.1\% of observations on average, with the average weight among downweighted observations ranging from 0.37 to 0.50 across metrics. The estimates agree with the MLE in periods without extreme observations (e.g., Metric A in 2023-H1 and 2023-H2), confirming that BIMLE does not distort estimation under normal conditions.

Order-of-magnitude swings in estimated prior parameters across adjacent half-year periods are difficult to rationalize as genuine changes in the distribution of test effects. Some evolution is expected as the testing portfolio changes, but a prior standard deviation that jumps from 0.016\% to 0.900\% more plausibly reflects the influence of a handful of extreme observations than a fiftyfold shift in underlying variability.

Appendix \ref{sec:appendix_supplemental} reports analogous results for $\hat\mu$. Consistent with the score for $\mu$ being linear in $z_i$, all three estimators are more stable when estimating the prior mean. The exception is 2025-H2, where extreme observations pull the MLE and MAP estimates while BIMLE is relatively stable.

\section{Conclusion}

This paper proposes a bounded influence maximum likelihood estimator (BIMLE) for estimating prior parameters in the normal-normal empirical Bayes model. The estimator is motivated by a simple observation: estimates of the variance of true impacts $\sigma^2$ can be highly sensitive to a small number of observations. By bounding the contribution of individual observations to the likelihood, BIMLE provides a principled and practical mechanism for reducing this sensitivity while retaining the full sample of tests in estimation. 

The approach is not without tradeoffs. Extreme observations may contain legitimate information regarding the tails of the treatment effect distribution, and any bounded influence procedure necessarily limits the extent to which such observations affect the estimates. Consequently, BIMLE should not be interpreted as recovering the ``true'' variance of treatment effects in all settings. Rather, it represents a deliberate tradeoff between responsiveness to extreme observations and stability of the resulting estimates.

In our view, this tradeoff is often desirable in production experimentation systems. These systems routinely process hundreds if not thousands of tests, making it difficult to manually determine whether extreme observations reflect genuine treatment effects or data quality issues. In such settings, allowing a small number of observations to exert unbounded influence on estimates of the prior variance can lead to unstable shrinkage rules and launch decisions across the firm. The empirical results suggest that substantial improvements in stability can be achieved while downweighting only a small fraction of observations and without excluding any tests from estimation.

\bibliographystyle{aer}
\bibliography{bibliography}

\clearpage 

\appendix

\section{Supplemental Results}
\label{sec:appendix_supplemental}

\vspace{-0.5pc}

\begin{table}[ht]
\centering
\small
\caption{Estimates of $\hat\mu \times 10{,}000$ by half-year period}
\label{tab:empirical_mu}
\begin{subtable}[t]{0.48\textwidth}
\centering
\caption{Metric A}
{\scriptsize
\renewcommand{\arraystretch}{0.85}
\begin{tabular}{lrrr}
\toprule
 & \multicolumn{1}{c}{MLE} & \multicolumn{1}{c}{MAP} & \multicolumn{1}{c}{BIMLE} \\
\midrule
2023-H1 & \shortstack{0.10\\{}[-0.07, 0.28]} & \shortstack{0.10\\{}[-0.07, 0.28]} & \shortstack{0.10\\{}[-0.07, 0.28]} \\
2023-H2 & \shortstack{0.02\\{}[-0.12, 0.16]} & \shortstack{0.02\\{}[-0.12, 0.16]} & \shortstack{0.02\\{}[-0.12, 0.16]} \\
2024-H1 & \shortstack{0.21\\{}[0.05, 0.40]} & \shortstack{0.20\\{}[0.05, 0.40]} & \shortstack{0.21\\{}[0.05, 0.40]} \\
2024-H2 & \shortstack{-0.09\\{}[-1.43, 0.08]} & \shortstack{-0.09\\{}[-1.38, 0.08]} & \shortstack{0.06\\{}[-0.10, 0.22]} \\
2025-H1 & \shortstack{0.09\\{}[-0.06, 0.22]} & \shortstack{0.09\\{}[-0.06, 0.22]} & \shortstack{0.08\\{}[-0.07, 0.22]} \\
2025-H2 & \shortstack{-0.98\\{}[-1.92, -0.01]} & \shortstack{-0.98\\{}[-1.91, -0.01]} & \shortstack{0.22\\{}[0.07, 0.38]} \\
\midrule
CV & 3.66 & 3.67 & 0.64 \\
DW & -- & -- & 0.02\% \\
$w | $ DW & -- & -- & 0.50 \\
\bottomrule
\end{tabular}
}
\end{subtable}
\hfill
\begin{subtable}[t]{0.48\textwidth}
\centering
\caption{Metric B}
{\scriptsize
\renewcommand{\arraystretch}{0.85}
\begin{tabular}{lrrr}
\toprule
 & \multicolumn{1}{c}{MLE} & \multicolumn{1}{c}{MAP} & \multicolumn{1}{c}{BIMLE} \\
\midrule
2023-H1 & \shortstack{0.12\\{}[-0.09, 0.32]} & \shortstack{0.12\\{}[-0.09, 0.31]} & \shortstack{0.12\\{}[-0.09, 0.32]} \\
2023-H2 & \shortstack{-0.03\\{}[-0.24, 0.20]} & \shortstack{-0.03\\{}[-0.24, 0.20]} & \shortstack{-0.03\\{}[-0.24, 0.20]} \\
2024-H1 & \shortstack{0.11\\{}[-0.09, 0.34]} & \shortstack{0.11\\{}[-0.09, 0.34]} & \shortstack{0.11\\{}[-0.09, 0.33]} \\
2024-H2 & \shortstack{0.04\\{}[-0.17, 0.28]} & \shortstack{0.04\\{}[-0.17, 0.28]} & \shortstack{0.05\\{}[-0.16, 0.29]} \\
2025-H1 & \shortstack{0.11\\{}[-0.07, 0.32]} & \shortstack{0.11\\{}[-0.07, 0.32]} & \shortstack{0.11\\{}[-0.07, 0.32]} \\
2025-H2 & \shortstack{-1.28\\{}[-2.32, 0.15]} & \shortstack{-1.28\\{}[-2.30, 0.15]} & \shortstack{0.16\\{}[-0.06, 0.35]} \\
\midrule
CV & 3.22 & 3.23 & 0.72 \\
DW & -- & -- & 0.01\% \\
$w | $ DW & -- & -- & 0.41 \\
\bottomrule
\end{tabular}
}
\end{subtable}

\begin{subtable}[t]{0.48\textwidth}
\centering
\caption{Metric C}
{\scriptsize
\renewcommand{\arraystretch}{0.85}
\begin{tabular}{lrrr}
\toprule
 & \multicolumn{1}{c}{MLE} & \multicolumn{1}{c}{MAP} & \multicolumn{1}{c}{BIMLE} \\
\midrule
2023-H1 & \shortstack{1.09\\{}[0.68, 1.64]} & \shortstack{1.09\\{}[0.68, 1.64]} & \shortstack{0.87\\{}[0.52, 1.27]} \\
2023-H2 & \shortstack{2.37\\{}[1.18, 3.59]} & \shortstack{2.37\\{}[1.18, 3.58]} & \shortstack{1.13\\{}[0.74, 1.47]} \\
2024-H1 & \shortstack{1.34\\{}[0.11, 2.49]} & \shortstack{1.33\\{}[0.11, 2.48]} & \shortstack{1.85\\{}[1.24, 2.48]} \\
2024-H2 & \shortstack{-1.12\\{}[-2.27, -0.14]} & \shortstack{-1.11\\{}[-2.26, -0.14]} & \shortstack{0.93\\{}[0.45, 1.36]} \\
2025-H1 & \shortstack{0.43\\{}[-0.08, 0.87]} & \shortstack{0.43\\{}[-0.08, 0.87]} & \shortstack{0.73\\{}[0.42, 1.03]} \\
2025-H2 & \shortstack{-1.41\\{}[-2.81, 0.03]} & \shortstack{-1.40\\{}[-2.78, 0.03]} & \shortstack{1.29\\{}[0.92, 1.65]} \\
\midrule
CV & 2.98 & 2.96 & 0.33 \\
DW & -- & -- & 0.07\% \\
$w | $ DW & -- & -- & 0.40 \\
\bottomrule
\end{tabular}
}
\end{subtable}
\hfill
\begin{subtable}[t]{0.48\textwidth}
\centering
\caption{Metric D}
{\scriptsize
\renewcommand{\arraystretch}{0.85}
\begin{tabular}{lrrr}
\toprule
 & \multicolumn{1}{c}{MLE} & \multicolumn{1}{c}{MAP} & \multicolumn{1}{c}{BIMLE} \\
\midrule
2023-H1 & \shortstack{0.38\\{}[-0.05, 0.63]} & \shortstack{0.38\\{}[-0.05, 0.63]} & \shortstack{0.41\\{}[0.20, 0.63]} \\
2023-H2 & \shortstack{0.20\\{}[-0.05, 0.73]} & \shortstack{0.20\\{}[-0.05, 0.73]} & \shortstack{0.31\\{}[0.10, 0.52]} \\
2024-H1 & \shortstack{0.27\\{}[-0.13, 0.49]} & \shortstack{0.27\\{}[-0.13, 0.49]} & \shortstack{0.36\\{}[0.12, 0.58]} \\
2024-H2 & \shortstack{-1.46\\{}[-3.11, 0.02]} & \shortstack{-1.45\\{}[-3.11, 0.02]} & \shortstack{0.22\\{}[-0.00, 0.46]} \\
2025-H1 & \shortstack{0.34\\{}[0.16, 0.52]} & \shortstack{0.34\\{}[0.16, 0.52]} & \shortstack{0.36\\{}[0.17, 0.54]} \\
2025-H2 & \shortstack{-1.94\\{}[-3.09, -0.32]} & \shortstack{-1.93\\{}[-3.06, -0.32]} & \shortstack{0.50\\{}[0.30, 0.70]} \\
\midrule
CV & 2.59 & 2.59 & 0.24 \\
DW & -- & -- & 0.04\% \\
$w | $ DW & -- & -- & 0.37 \\
\bottomrule
\end{tabular}
}
\end{subtable}
\vspace{0.1pc}
{\scriptsize
\begin{flushleft}
Notes: BIMLE uses $b = 15$. MAP uses $\mu \sim N(0, 10^{-6})$ and $\log(\sigma^2) \sim N(-16, 2^2)$. Brackets show 95\% confidence intervals from a nonparametric bootstrap (200 resamples of the tests within each half-year window). CV is the coefficient of variation across half-year periods. DW is the average percent of observations receiving weight less than 1. $w | $ DW is the average weight among downweighted observations.
\end{flushleft}}
\end{table}

\vspace{-1.5pc}

\begin{table}[ht]
\centering
\footnotesize
\caption{BIMLE estimates of $\hat\sigma \times 100$: sensitivity to $b$}
\label{tab:b_sensitivity}
\vspace{0.5pc}
\begin{tabular}{lrrrrrrrrr}
\toprule
 & \multicolumn{3}{c}{$b = 10$} & \multicolumn{3}{c}{$b = 15$} & \multicolumn{3}{c}{$b = 20$} \\
\cmidrule(lr){2-4} \cmidrule(lr){5-7} \cmidrule(lr){8-10}
 & Mean & CV & DW & Mean & CV & DW & Mean & CV & DW \\
\midrule
Metric A & 0.019 & 0.16 & 0.04 & 0.021 & 0.15 & 0.02 & 0.022 & 0.19 & 0.02 \\
Metric B & 0.026 & 0.16 & 0.02 & 0.028 & 0.17 & 0.01 & 0.029 & 0.22 & 0.01 \\
Metric C & 0.206 & 0.14 & 0.17 & 0.261 & 0.23 & 0.07 & 0.291 & 0.24 & 0.04 \\
Metric D & 0.071 & 0.10 & 0.06 & 0.080 & 0.15 & 0.04 & 0.085 & 0.16 & 0.03 \\
\bottomrule
\end{tabular}
\vspace{0.1pc}
{\scriptsize
\begin{flushleft}
Notes: Mean and CV of $\hat\sigma \times 100$ across half-year periods. DW is the average percent of observations receiving weight less than 1.
\end{flushleft}}
\end{table}

\vspace{-1.5pc}

\begin{table}[H]
\centering
\footnotesize
\caption{BIMLE estimates of $\hat\mu \times 10{,}000$: sensitivity to $b$}
\label{tab:b_sensitivity_mu}
\vspace{0.5pc}
\begin{tabular}{lrrrrrrrrr}
\toprule
 & \multicolumn{3}{c}{$b = 10$} & \multicolumn{3}{c}{$b = 15$} & \multicolumn{3}{c}{$b = 20$} \\
\cmidrule(lr){2-4} \cmidrule(lr){5-7} \cmidrule(lr){8-10}
 & Mean & CV & DW & Mean & CV & DW & Mean & CV & DW \\
\midrule
Metric A & 0.12 & 0.65 & 0.04 & 0.12 & 0.64 & 0.02 & 0.11 & 0.66 & 0.02 \\
Metric B & 0.09 & 0.71 & 0.02 & 0.08 & 0.72 & 0.01 & 0.08 & 0.71 & 0.01 \\
Metric C & 0.91 & 0.24 & 0.17 & 1.13 & 0.33 & 0.07 & 1.17 & 0.37 & 0.04 \\
Metric D & 0.38 & 0.21 & 0.06 & 0.36 & 0.24 & 0.04 & 0.34 & 0.26 & 0.03 \\
\bottomrule
\end{tabular}
\vspace{0.1pc}
{\scriptsize
\begin{flushleft}
Notes: Mean and CV of $\hat\mu \times 10{,}000$ across half-year periods. DW is the average percent of observations receiving weight less than 1.
\end{flushleft}}
\end{table}

\end{document}